\magnification=1200
\input iopppt.modifie
\input epsf
\def\received#1{\insertspace 
     \parindent=\secindent\ifppt\textfonts\else\smallfonts\fi 
     \hang{#1}\rm } 
\def\figure#1{\global\advance\figno by 1\gdef\labeltype{\figlabel}%
   {\parindent=\secindent\smallfonts\hang 
    {\bf Figure \ifappendix\applett\fi\the\figno.} \rm #1\par}} 
\headline={\ifodd\pageno{\ifnum\pageno=\firstpage\titlehead
   \else\rrhead\fi}\else\lrhead\fi}

\def\rrhead{\textfonts\hskip\secindent\it 
    \shorttitle\hfill\rm L\folio} 
\def\lrhead{\textfonts\hbox to\secindent{\rm L\folio\hss}%
    \it\aunames\hss} 
\footline={\ifnum\pageno=\firstpage
\hfil\textfonts\rm L\folio\fi}   
\def\titlehead{\smallfonts J. Phys. A: Math. Gen.  {\bf 24} (1991) 
L1229--L1234\hfil} 

\firstpage=1229
\pageno=1229

\jnlstyle
\jl{1}
\overfullrule=0pt

\letter{Critical behaviour in parabolic geometries}[Letter to the Editor]

\author{Ingo Peschel\footnote{\dag}{Permanent address: Fachbereich
 Physik, Freie Universit\"at Berlin, Arnimallee 14\hfill\break
W--1000 Berlin 33,
 Federal Republic of Germany}, Lo\"\ii c Turban and Ferenc Igl\'oi\footnote{\ddag}{Permanent address: Central Research Institute for 
Physics, H--1525 Budapest, Hungary}}[Letter to the Editor]
 
\address{Laboratoire de Physique du Solide\footnote{\S}{URA CNRS no155}, 
Universit\'e de Nancy I, BP239\hfill\break F--54506 
Vand\oe uvre l\`es Nancy Cedex, France}
\received{Received 15 July 1991}
\abs
We study two--dimensional systems with boundary curves described 
by power laws. Using conformal mappings we obtain the 
correlations at the bulk critical point. Three different classes of
behaviour are found and explained by scaling arguments which also 
apply to higher dimensions.
For an Ising system of parabolic shape the behaviour of the order
at the tip is also found.
\endabs

\vglue1.5cm

The shape of a system undergoing a second order phase transition
can have a strong influence on its critical behaviour. This is 
shown by the results for edges (in~{\smallfonts3D}) or corners 
(in~{\smallfonts2D}). The local critical exponents are then
continuous functions of the corresponding angle~[1--6]. But this
striking feature also raises the question which  property of the
boundary actually causes it and what would be obtained for other
shapes. For the critical behaviour long range effects are essential
and thus a simple rounding of the corner will not matter~[7]. We
therefore study here shapes which differ from the corner geometry in
the large: the boundary curves are described by power laws and  do
not have asymptotes. The prototype is the parabola. We use  conformal
mappings to obtain the critical correlation functions for  various
two--dimensional geometries. When the system forms the interior of a
general parabolic figure, we find a new unusual form of the critical 
behaviour. When the system forms the exterior, on the other hand, one
recovers the behaviour of a system with either a straight surface or
a cut. These results can be understood from the way the boundary
curves behave under renormalization. A similar classification will 
therefore hold in three dimensions. Our results at the critical point
are complemented by a calculation of the tip magnetization for an
Ising model of parabolic shape which also shows unusual features.
{\par\begingroup\parindent=0pt\medskip \epsfxsize=13truecm
\topinsert
\centerline{\epsfbox{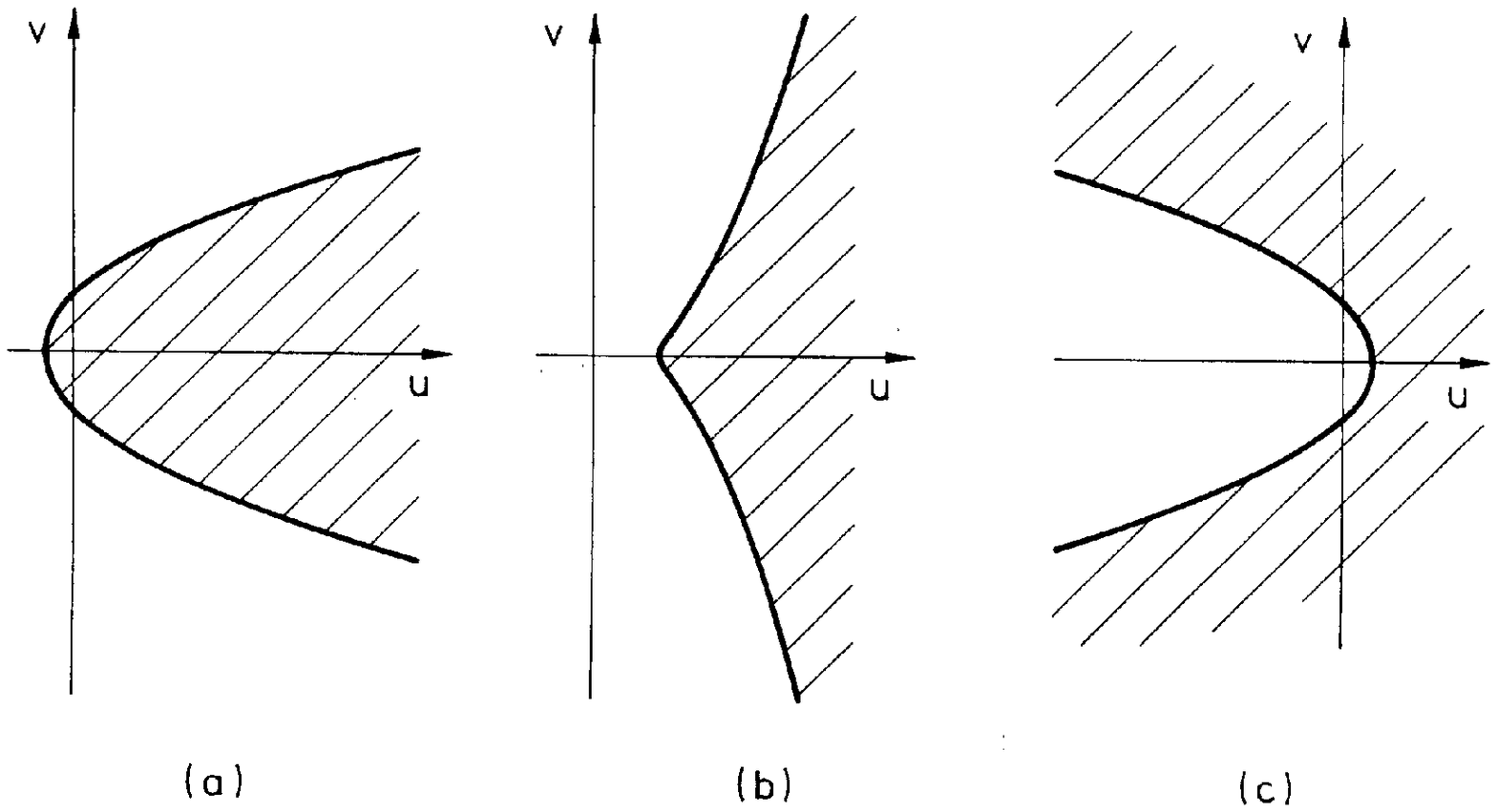}}
\smallskip
\figure{The three types of geometries considered in the text.} 
\endinsert 
\endgroup
\par}
Consider first a system with free boundaries in the form of a simple 
parabola $v^2=2pu+p^2$ in the plane $w=u+\i v$ as in figure (1a). It 
can be related to the half--plane $z=x+\i y$, $y>0$ by the conformal 
map
$$
z=\i \cosh\left(\pi\sqrt{w\over2p}\right).\eqno(1)
$$
At criticality, the correlation function in the half--plane has the 
form~[3]
$$
G(z_1,z_2)=(y_1y_2)^{-x}\psi(\omega)\eqno(2)
$$
where the scaling function depends on the variable $\omega=4y_1y_2
/\vert z_1-z_2\vert^2$ and has the asymptotic form $\psi(\omega)\sim
\omega^{x_1}$ for small $\omega$. Here $x$ and $x_1$ denote the bulk
and surface scaling dimensions, respectively, of the operators in $G$.
Using the standard transformation~[8], one finds in the parabolic 
geometry
$$
G(w_1,w_2)=\lambda_1^x\lambda_2^x
(\cosh\zeta_1\cosh\zeta_2\cos\eta_1\cos\eta_2)^{-x}\psi(\omega)
\eqno(3)
$$
where the rescaling factors are
$$
\lambda_i={\pi^2\over 4p}\left({\sinh^2\zeta_i+\sin^2\eta_i\over
\zeta_i^2+\eta_i^2}\right)^{1/2}\eqno(4)
$$
and we have used parabolic coordinates $\pi \sqrt{w\over 2p}=
\zeta +\i \eta$. The system is then characterized by $0\leq \eta 
\leq \pi /2$. If $w_1$, $w_2$ lie on the positive $u$-axis, the 
variable $\omega$ is given by $\omega=4\cosh\zeta_1\cosh\zeta_2
/(\cosh\zeta_1-\cosh\zeta_2)^2$. One then finds for $\zeta_1$ 
fixed and $\zeta_2\gg 1$ $(u_2\gg p)$
$$
G(w_1,w_2)=A(u_1){1\over u_2^{x/2}}\exp\left(-\pi x_1
\sqrt{u_2\over 2p}\right).\eqno(5)
$$
This is not a simple power law in $u_2$ as one finds for a corner, 
which clearly shows the difference between the two cases. It also 
differs from the result in a strip. However, if both $u_1,u_2
\gg p$ and $\sqrt{u_2}-\sqrt{u_1}\gg\sqrt p$, it can be written as
$$
G(w_1,w_2)=\left({\pi^2\over L(u_1)L(u_2)}\right)^x\exp 
\left[-2\pi x_1\left({u_2\over L(u_2)}-{u_1\over L(u_1)}\right)
\right]\eqno(6)
$$
with $L(u)=2\sqrt{2pu}$ being the width of the system at position 
$u$. If in addition $u_2-u_1\ll (u_2+u_1)/2=u$, an expansion around 
$u$ in (6) gives back the strip result~[8]. In this sense, the 
parabola can be considered as a strip of varying width.

If one fixes the boundary variables one can also discuss the order 
parameter profile. A transformation as in (3) then gives, for $w=
u>0$
$$
\langle\phi (u)\rangle=A\left({\pi^2\tanh \zeta\over 4p\zeta}\right)^x.\eqno(7)
$$
Thus $\langle\phi\rangle\sim u^{-x/2}$ for $u\gg p$ and there is no 
exponential factor in this case. The exponent, however, is still
different from its value $x$ found for a corner.

One can easily generalize the treatment to boundaries which have 
the asymptotic form 
$$
v=\pm Cu^\alpha \qquad \alpha <1.\eqno(8)
$$
One merely has to shift the parabola into the right half--plane
and then to distort it. This amount to the replacement
$$
{w\over 2p}\rightarrow \left({w\over 2p}\right)^{2(1-\alpha )}-
{1\over 2}\eqno(9)
$$
in equation (1). The quantity $C$ is then $C=(2p)^{1-\alpha }/2
(1-\alpha )$. This changes the result (5) for $G$ into
$$
G(w_1,w_2)=A(u_1){1\over u_2^{\alpha x}}\exp \left[{-\pi x_1\over 
2C(1-\alpha )}u_2^{1-\alpha }\right].\eqno(10)
$$
The functional form of $G$ thus varies continuously with the 
parameter $\alpha $ describing the boundary shape. The more this 
shape 
approaches the corner geometry $(\alpha \rightarrow 1)$, the slower 
the exponential falloff becomes. On the other hand, for $\alpha =0$
the system forms a half--strip and one recovers the corresponding
simple exponential decay~[8,9]. For $\alpha <0$, it has a
spoon--like  shape and the decay becomes very rapid in the narrow
region.

We now turn to a system in the shape of figure (1b). Here the
boundary is curved towards the outside, so that $\alpha >1$ in 
equation (8). To relate it to the upper $z$-plane one has to use a 
different mapping, namely
$$
z=\i \left[w^s-\left({p\over 2}\right)^s\right]^{1/s}\eqno(11)
$$
where $s=1-1/\alpha $. Asymptotically, one now has $z=\i w$ and the 
rescaling factor $\vert \d w/\d z\vert$ appearing in the 
transformation of $G$ becomes one. Therefore, for $w_1,w_2$ on the 
axis, with $u_1$ fixed and $u_2 \rightarrow \infty$, one always 
obtains the result of the half--plane
$$
G(w_1,w_2)=A(u_1){1\over u_2^{x+x_1}}.\eqno(12)
$$
In this sense, this type of boundary is equivalent to a straight 
surface.

Finally, a system with a cut--out portion in the form of equation
(8) as in figure (1c), can be obtained via a mapping
$$ 
z=\i \left[w^s-\left({p\over 2}\right)^s\right]^{1/2s}\eqno(13)
$$
where now $s=1-\alpha $ and $\alpha <1$ again. Asymptotically, the 
relation therefore is, for all $\alpha $, $z=\i \sqrt w$. But with 
such a transformation one maps the $z$-half--plane onto the
$w$-plane with a cut~[2--4].The correlation function therefore is 
asymptotically
$$
G(w_1,w_2)=A(u_1){1\over u_2^{x+x_2}}\eqno(14)
$$
with the corner exponent $x_2={1\over 2}x_1$ corresponding to 
the cut.

The preceding results can be understood if one considers the 
behaviour of the boundary curve, equation (8), under a change 
of scale in a renormalization procedure. With $u'=u/b$, $v'=v/b$
it becomes
$$ 
v'=\pm b^{\alpha -1}C(u')^\alpha \eqno(15)
$$
or
$$
C'=b^{\alpha -1}C\eqno(16)
$$
Thus, for $\alpha >1$, $C$ grows under renormalization and 
the boundary curve approaches a straight line. For $\alpha =1$, $C$
is invariant and thus a marginal variable. This explains the 
particular role of a corner formed by two straight lines. For
$\alpha <1$, $C$ decreases and the system approaches either a cut
geometry or a one--dimensional line geometry. In the latter case,
however, one has a non--ordering system and 
this causes the particular features of the parabolic geometry.

According to equation (16), $1/C$ may be considered as a scaling 
field with dimension $1-\alpha$ (like $1/L$ in finite--size scaling).
It vanishes at the half--plane fixed point. One may therefore write 
the following scaling ansatz for the correlations along the $u$-axis
$$
G\left(u_1,u_2,{1\over C}\right)=b^{-2x}G\left({u_1\over b},
{u_2\over b},{b^{1-\alpha}\over C}\right).\eqno(17)
$$
With $b=C^{1/1-\alpha}$, one gets
$$
G\left(u_1,u_2,{1\over C}\right)=C^{-2x/1-\alpha}
g\left({u_1\over L(u_1)},{u_2\over L(u_2)}\right)\eqno(18)
$$
where $L(u)=2Cu^\alpha$ is the width of the system at $u$. Equation
(6) can thereby be generalized to any value of $\alpha<1$ with 
the scaling function given by
$$
g(a_1,a_2)\sim (a_1a_2)^{-x\alpha/1-\alpha}
\exp\left[-{\pi x_1\over 1-\alpha}(a_2-a_1)\right]\eqno(19)
$$
when $a_1,a_2\gg 1$ and $a_2-a_1\gg 1$. This can also be verified 
explicitly. The scaling behaviour of the order parameter profile 
is obtained in the same way and reads
$$
\langle\phi(u)\rangle=L(u)^{-x}f\left({u\over L(u)}\right)
\eqno(20)
$$
where, according to equation (7), $\lim_{a\rightarrow\infty}f(a)
=O(1)$.

Relations like (18) and (20) correspond to a local formulation 
of finite--size scaling. One may also notice that all 
these scaling considerations still apply in higher dimensions.

Finally, let us address briefly the ordered state. A priori, it is
not obvious that a system in the shape of figure (1a) will order
at all. We have therefore studied an Ising model with parabolic
shape $v=\pm C\sqrt u$, using the corner transfer matrix
technique~[10]. This means that one considers the transfer matrix
connecting  the spins at the upper and lower boundaries with fixed
boundary  condition on the right end of the system. Assuming a square
lattice  in the Hamiltonian limit~[11] one then is lead to study the
following operator (describing an inhomogeneous transverse Ising
chain) $$ H=-C\left[\sum_{n=1}^{N-1}\sqrt n\ \sigma_n^z+\lambda 
\sum_{n=0}^{N-1}\sqrt{n+1}\ \sigma_n^x\sigma_{n+1}^x\right]\eqno(21)
$$
where $\lambda^{-1}$ measures the temperature and $N$ is the size 
of the system along the axis. The coefficients reflect, in
the sense of a continuum limit, the number of vertical and of 
horizontal bonds at position $n$, respectively. The transverse field 
vanishes at $n=0$ due to the absence of vertical bonds for the first 
spin and at $n=N$ as a consequence of the boundary condition.
The operator can be diagonalized in 
terms of fermions. The single--particle excitation energies 
$\epsilon_\nu=2C\omega_\nu$ then follow from
$$
n\lambda\psi_{n-1}^\nu +n(\lambda^2+1)\psi_n^\nu +n\lambda
\psi_{n+1}^\nu =\omega_\nu^2\psi_n^\nu \eqno(22)
$$
with appropriate boundary conditions at $n=0,\ N$. This system
of equations is similar to one studied previously in 
a related context~[12] and, as there, can be solved with Gottlieb
polynomials. In the limit $N\rightarrow\infty$ one finds, for 
$\lambda>1$ 
$$
\omega_\nu=\sqrt{(\lambda^2-1)\nu},\qquad\nu=1,2,3\cdots\eqno(23)
$$
Identifying the boundaries, the magnetization at the 
tip of a system which is isotropic at the critical point, 
is given by~[10,11]
$$
m_0=\langle\sigma_0^x\rangle=\prod_\nu \tanh\left({\epsilon_\nu\over 2}\right).
\eqno(24)
$$
Evaluating this near the critical point $(\lambda\geq 1)$ leads to
$$
m_0\sim \exp \left[{-a\over C^2(\lambda -1)}\right]\eqno(25)
$$
where $a=7\zeta(3)/16\simeq 0.526$. Thus there is order, but it 
vanishes exponentially fast at the critical point. This reflects 
the difficulty to maintain it in such a geometry. We note that
the argument in the exponential can be expressed as the ratio
$\xi /p$ where $\xi\sim (\lambda-1)^{-1}\sim t^{-\nu}$ is the bulk 
correlation length.

The behaviour of the tip magnetization may also be deduced from 
scaling considerations. The magnetization at position $u$ along 
the axis satisfies
$$
m\left(t,u,{1\over C}\right)=b^{-x}m\left(b^{1/\nu}t,{u\over b}
,{b^{1-\alpha}\over C}\right)\eqno(26)
$$
which, with $b=t^{-\nu}$, leads to
$$
m\left(t,u,{1\over C}\right)=t^\beta f\left({u\over t^{-\nu}},
{t^{-\nu(1-\alpha)}\over C}\right)\eqno(27)
$$
where $\beta$ and $\nu$ are bulk exponents. 

One may even go further assuming that, at the tip, the leading 
contribution to the magnetization which is induced by the bulk at 
a distance $D\sim (\xi/C)^{1/\alpha}$ (where the width of the 
system is of the order of the bulk correlation length), decays 
with $D$ like the correlation function in (10) when $u_1\rightarrow 
0$ and $u_2=D\gg C^{1/1-\alpha}$. Then
$$
m\left(t,0,{1\over C}\right)=m_0\sim \exp\left(-a{D^{1-\alpha}
\over C(1-\alpha)}\right)\eqno(28)
$$
and the temperature dependence of the tip magnetization follows
$$
m_0\sim \exp\left[-a\left({t^{-\nu(1-\alpha)}
\over C(1-\alpha)}\right)^{1/\alpha}\right]\eqno(29)
$$
in agreement with (24) for the Ising parabola with $\alpha=
{1\over 2}$ and $\nu=1$. 

One should mention that the results (10) for the correlation 
function and (29) for the order parameter are quite similar to
those obtained for an Ising model with bond strengths decreasing
towards a free surface as $K(n)=K(\infty)(1-A/n^y)$, $y<1$~[13--15]
with the correspondences $y\leftrightarrow \alpha$, 
$A\leftrightarrow C$. This can be understood qualitatively since  in
both cases the surface order near the critical point can only be
maintained through the action of the far--away bulk portion of the
system. 

Finally, for an anisotropic system with correlation length 
exponents $\nu_\parallel$ (along the $u$-axis) $\not =\nu_\perp$,
the scaling dimension of $1/C$ is changed into $1-\alpha\nu_
\parallel/\nu_\perp$. Therefore the perturbation to the half--plane 
geometry then becomes relevant when $\alpha<\nu_\perp/\nu_\parallel$.

\ack
IP thanks the Laboratoire de Physique du Solide for the 
hospitality extended to him in Nancy. FI gratefully acknowledges
the financial support of the Minist\`ere Fran\c cais des Affaires
Etrang\`eres through a research grant.

\references
\numrefjl{[1]}{Cardy J L 1983}{\JPA}{16}{3617}
\numrefjl{[2]}{Barber M N, Peschel I and Pearce PA 1984}
{J. Stat. Phys.}{37}{497}
\numrefjl{[3]}{Cardy J L 1984}{\NP\ \rm B}{240}{514}
\numrefjl{[4]}{Bariev R Z 1986}{Teor. Mat. Fiz.}{69}{149}
\numrefjl{[5]}{Kaiser C and Peschel I 1989}{J. Stat. Phys.}{54}{567}
\numrefjl{[6]}{Davies B and Peschel I 1991}{\JPA}{24}{1293}
\numrefbk{[7]}{Diehl H W 1986}{Phase Transitions and Critical 
Phenomena}{vol 10 ed. C Domb and J L Lebowitz (London: Academic Press) 
p 241}
\numrefbk{[8]}{Cardy J L 1987}{Phase Transitions and Critical 
Phenomena}{vol 11 ed. C Domb and J L Lebowitz (London: Academic Press)
p 68}
\numrefjl{[9]}{Cardy J L 1984}{\JPA}{17}{L385}
\numrefbk{[10]}{Baxter R J 1982}{Exactly Solved Models in Statistical 
Mechanics}{(London: Academic Press) p~363}
\numrefjl{[11]}{Peschel I and Truong T T 1987}{\ZP\ \rm B}{69}{385}
\numrefjl{[12]}{Truong T T and Peschel I 1990}{Int. J. Mod. Phys.}
{4}{895}
\numrefjl{[13]}{Hilhorst H J and van Leeuwen J M J 1981}{\PRL}{47}{1188}
\numrefjl{[14]}{Burkhardt T W, Guim I, Hilhorst H J and van Leeuwen 
J M J 1984}{\PR\ \rm B}{30}{1486}
\numrefjl{[15]}{Peschel I 1984}{\PR\ \rm B}{30}{6783}

\vfill\eject
\bye